\documentclass[onecollarge]{svjour2}       
\smartqed  
\usepackage{graphicx}
\journalname{Journal of Superconductivity}
\begin{document}

\title{Vortex state microwave resistivity in Tl-2212 thin films.
}

\titlerunning{Vortex state microwave resistivity in Tl-2212}        

\author{N. Pompeo         \and
	S. Sarti \and
        R. Marcon \and 
	H. Schneidewind \and
	E. Silva 
}

\authorrunning{N. Pompeo at al.} 

\institute{
	   N. Pompeo \at Dipartimento di Fisica "E. Amaldi" and
	   Unit\`a CNISM, Universit\`a di Roma Tre, Via della Vasca
	   Navale 84, 00146 Roma, Italy
	   \and
	   S. Sarti \at Dipartimento di Fisica and Unit\`a CNISM,
	   Universit\`a "La Sapienza", 00185 Roma, Italy
           \and
	   R. Marcon \at Dipartimento di Fisica "E. Amaldi" and
	   Unit\`a CNISM, Universit\`a di Roma Tre, Via della Vasca
	   Navale 84, 00146 Roma, Italy
	   \and 
	   H. Schneidewind \at Institute for Physical High
	   Technology Jena, P.O.B.100239, D-07702 Jena, Germany
	   \and 
	   E. Silva \at Dipartimento di Fisica "E. Amaldi" and
	   Unit\`a CNISM, Universit\`a di Roma Tre, Via della Vasca
	   Navale 84, 00146 Roma, Italy \\
              Tel.: +39-06-55177205\\
              Fax: +39-06-5579303\\
              \email{silva@fis.uniroma3.it} 
}

\date{Received: date / Accepted: date}

\maketitle

\begin{abstract}
We present measurements of the field induced changes in the 47 GHz
complex resistivity, $\Delta \tilde \rho(H,T)$, in
Tl$_{2}$Ba$_{2}$CaCu$_{2}$O$_{8+x}$ (TBCCO) thin films with $T_{c}\simeq$ 105
K, prepared on CeO$_{2}$ buffered sapphire substrates.  At low fields
($\mu_{0}H<$10 mT) a very small irreversible feature is present,
suggesting a little role of intergranular phenomena.  Above that
level $\Delta \tilde \rho(H,T)$
exhibits a superlinear dependence with the field, as opposed to the
expected (at high frequencies) quasilinear behaviour.  We observe a
crossover between predominantly imaginary to predominantly real
(dissipative) response with increasing temperature and/or field.  In
addition, we find the clear scaling property $\Delta \tilde
\rho(H,T)=\Delta \tilde \rho[H/H^{*}(T)]$, where the scaling field
$H^{*}(T)$ maps closely the melting field measured in single crystals.
We discuss our microwave results in terms of loss of flux lines
rigidity.
%
\end{abstract}

\section{Introduction}
\label{intro}
The microwave response of type-II superconductors in the vortex state
is determined by a rich variety of phenomena.  In principle, the
response is dictated by intrinsic processes, such as the motion of
Abrikosov flux lines, described by the vortex motion complex
resistivity $\tilde\rho_{vm}=\rho_{1}+\rm{i}\rho_{2}$, and charge 
conduction, described by the complex
conductivity $\tilde\sigma$, and by many extrinsic processes, related to the granularity
of real samples.  Focusing for the moment on intrinsic processes, when
the temperature is not too close to $T_{c}$ the motion of flux lines
(vortices) is thought to dominate the vortex state electrodynamic
response up and beyond the microwave spectrum.\\
A general approach to the field, frequency and temperature dependence
of the vortex motion (and its interplay with the response of the
condensate) is a formidable task, that remains to be fully developed.
Even the frequency dependence of vortex motion alone is still a debated
topic.  Nevertheless, when flux lines can be treated as rigid rods, or
as fully decoupled pancakes, it is believed that at high enough
frequencies the electrodynamics is essentially dictated by
single-vortex response.  This is due to the very small amplitude of
the oscillations induced by the alternate current at high frequencies.
In high-$T_{c}$ superconductors (HTS) at typical microwave frequencies
($\nu\sim$10 GHz, where $\nu$ is the microwave frequency) the
oscillations are estimated to be well below 1 nm \cite{tomasch}.  In
the single-vortex limit, the vortex motion gives rise to a microwave
response determined by the vortex viscosity $\eta$, which takes into
account the dissipation due to superfluid/quasiparticle conversion
during vortex motion \cite{golos} and is determined by the upper
critical field scale, and by the vortex pinning constant $\kappa_{p}$,
which takes into account the elastic response of the vortex.  The
depinning frequency $\nu_{p}=\kappa_{p}/2\pi\eta$ is also introduced,
and it marks the crossover between the pinning dominated (Campbell)
regime at $\nu\ll\nu_{p}$, where the response is predominantly
imaginary, to the viscous regime (flux flow) at $\nu\gg\nu_{p}$, where
the response is predominantly real and dissipative.  With just these
ingredients, one has the well-known Gittleman-Rosenblum model, where
the vortex motion complex resistivity is given by \cite{gittle}:
\begin{equation}
\label{GR}
\tilde\rho_{GR}(H,T)=
\frac{\Phi_{0}B}{\eta}
\frac{1+\mathrm{i}\frac{\nu_{p}}{\nu}}{1+\left(\frac{\nu_{p}}{\nu}\right)^{2}}
\end{equation}
where $\Phi_{0}=2.07 \times 10^{-15}$ Tm$^{-2}$ is the flux quantum.
When creep of vortices is allowed, Coffey and Clem \cite{cc} and Brandt 
\cite{brandt} extended the
model by accounting for a periodic potential and for a relaxational 
Labusch parameter, respectively. The resulting
vortex resistivity could be written in a very similar way as:
\begin{equation}
\label{ccb}
\tilde\rho_{CCB}(H,T)=
\frac{\Phi_{0}B}{\eta}
\frac{1+\epsilon\left(\frac{\nu_{0}}{\nu}\right)^{2}+\mathrm{i}(1-\epsilon)\frac{\nu_{0}}{\nu}}{1+\left(\frac{\nu_{0}}{\nu}\right)^{2}}
\end{equation}
where $\epsilon\approx e^{-\frac{U}{k_{B}T}}$ is the creep factor, 
$k_{B}$ is the Boltzmann constant and
$\nu_{0}\rightarrow\nu_{p}$ in the limit $\epsilon\rightarrow 0$.\\
These simple models have been effective in the description of the
frequency dependence of the microwave response in conventional
superconductors \cite{gittle} and in YBa$_{2}$Cu$_{3}$O$_{7-\delta}$
(YBCO) \cite{golos,revenaz,sartiPhC}, sometimes with the addition of a
field-independent \cite{marcon,tsuchiya} or field-dependent 
\cite{pompeoJPCS}
contribution of the superfluid.  It should also be mentioned that,
within these models, the elastic response can originate from both
pinning and elasticity of the vortex line, so that care should be
taken when discussing the parameters resulting from the analysis of
the data: extrinsic pinning phenomena as well as vortex line tension
can be in principle probed by the electromagnetic response.\\
It has to be noted that a crucial role is played by the measuring
frequency: by lowering the frequency below the characteristic
frequency $\nu_{0}$ (or $\nu_{p}$), vortices experience large drags
from their equilibrium positions and they can interact with each
other and with several potential wells.  In this case the nature and
distribution of pinning centers becomes crucial, and the various vortex
phases arising \cite{ffh,blatter} can have very different transport
properties.  The same validity of the single-vortex models can then be
questioned: as an experimental example, swept-frequency measurements
in YBCO have shown \cite{wu} that the microwave resistivity undergoes
a crossover between a collective (glassy) behaviour to the independent
vortex motion around a characteristic frequency of order of a few 
GHz.\\
Granularity is sometimes indicated as a possible dominant source for
the losses in the microwave response in superconducting films.
Manifestations of granularity include weak-links
dephasing \cite{giura}, Josephson fluxon (JF) dynamics
\cite{halbritter} and, as recently studied, Abrikosov - Josephson fluxon
(AJF) dynamics \cite{gurevich}. Weak-links dephasing is charaterized by
a very sharp increase of the dissipation at dc fields of order or less
than 20 mT, accompanied by a strong hysteresis \cite{giura}.
Josephson fluxon dynamics has been studied essentially in relation to
nonlinear effects, due to the short JF nucleation time.  In fact, it
has been reported \cite{tsindlekht} that in thin YBCO films nonlinear
and linear phenomena are decoupled, and that linear phenomena are
related to Abrikosov fluxons while JF play a role in the nonlinear
response only.  If however one assumes that a dc field has the same
effect as a microwave field, the qualitative properties of JF dynamics
would be barely distinguishable from conventional Abrikosov fluxons
dynamics, since an equation like Eq.\ref{GR} would
hold \cite{halbritter}, with the noticeable difference of a small
viscosity due to the insulating core.
Abrikosov-Josephson fluxons nucleate along small-angle grain
boundaries.  The ac response is expected to saturate at fields larger
than a characteristic field $H_{0}$, whose estimate spans orders of
magnitude \cite{gurevich} in the range 0.1-10 T. The explicit
expression for AJ fluxon motion ac resistance \cite{gurevich} yields an
initial magnetic-field increase as
$\sim\sqrt{\frac{H}{H_{0}}}$, with a subsequent saturation. \\
Up to now, the more intense experimental effort concerning the
microwave properties in the vortex state has been directed to the study
of YBCO, while other HTS did not receive the same attention.  In
particular, little is known about TBCCO, where the microwave response
has been studied essentially in connection with nonlinear properties
\cite{gaganidze} due to the potential interest for applications.  Aim
of this paper is to present a study of the microwave response of TBCCO
in the vortex state in moderate fields.  It will be shown that the
measured data, even if taken at the high microwave frequency of 47.7
GHz, do not exhibit free flux flow.  Instead, the data exhibit a giant
reactance in the intermediate field region, indicating very small flux
creep and strong elastic response for fields of order of a few kG. With
increasing temperature and field, we observe a crossover toward a
predominantly dissipative behaviour, with
$\rho_{1}>\rho_{2}$.  In the crossover and
dissipative regions the vortex complex resistivity undergoes a
field-dependent scaling: the curves of the complex resistivity
$\tilde\rho_{vm}(H,T)=\tilde \rho_{vm}[H/H^{*}(T)]$, where the temperature 
dependence of $H^{*}(T)$ reproduces
the temperature dependence of the melting field as measured in single
crystals, suggesting that the crossover from the elastic to the 
dissipative region is driven by the loss of the vortex rigidity.
\section{Experimental setup and samples}
\label{exp}
We measured the microwave response at high microwave frequencies 
at 47.7 GHz in TBCCO thin films, with a moderate magnetic 
field ($\mu_{0}H<$0.8 T) applied along the $c$ axis at temperatures 
above 59 K.
The 240 nm-thick films have been grown on
2'' diameter CeO$_2$ buffered R-plane sapphire substrates by
conventional two - step method.  The resulting films show excellent
(100) orientation
without any (111) components 
and excellent in-plane epitaxy \cite{schneidewind}.  The
full-width-half-maximum of the $\theta-2\theta$ rocking curve is
0.4$^{\circ}$.  The film under study had $T_{c}\simeq$104 K and
$J_{c}=0.5$ MAcm$^{-2}$ measured inductively.\\
We measured the magnetic field dependence of the complex resistivity
at 47.7 GHz by means of a sapphire dielectric resonator operating in
the TE$_{011}$ mode.  The sapphire rod was sandwiched between the
superconducting film and a copper plate.  With this configuration we
avoided averaging the response between two different films, at the
expense of a reduced sensitivity in the absolute value of the surface
impedance.  Nevertheless, the sensitivity was approximately two orders of
magnitude higher than with our preexisting metal cavity
\cite{silvaMST}.  The detailed description of the experimental setup
will be presented elsewhere \cite{pompeoJSsub}.  We checked that the
corrections for the film thickness \cite{sridhar} were negligible with 
respect to the thin-film approximation, so that (in
agreement with extensive simulations \cite{silvaSUSTthin}) the
measurements yield the microwave resistivity instead of the bulk
surface impedance.  We measured the changes of the $Q$ factor and of
the resonant frequency $f_{0}$ of the resonator, obtaining the changes
of the complex resistivity at a fixed temperature as:
$\Delta\tilde\rho(H,T)=\Delta\rho_{1}(H,T)+\mathrm{i}\Delta\rho_{2}(H,T)=
Gd\left[\left(\frac{1}{Q(H,T)}-\frac{1}{Q(0,T)}\right)-
2\mathrm{i}\frac{f_{0}(H,T)-f_{0}(0,T)}{f_{0}(0,T)}\right]$.  We stress
that in these measurements the variation of the effective impedance
does not depend on the calibration of the resonator, but only on the
geometrical factor $G\simeq$ 2000 $\Omega$ and on the film thickness
$d$, which act as a single scale factor.\\
Since field-dependent measurements are involved, a short discussion of
possible parasitic effects is in order.  In the present configuration
the microwave field probes a circular area, approximately of the size
of the diameter of the sapphire rod, $\sim$ 2 mm, centered on the
center of the film.  The edges of the film are not exposed to the
microwave field, so that our measurements do not probe vortex entry
and exit induced by microwave currents.  Due to the field distribution
of the excited mode, the peak microwave field $\mu_{0} H_{\mu w}\sim$
20 $\mathrm{\mu}$T is reached in an annular area
of $\sim$1 mm diameter centered on the center of the film.  Thus, for
all practical purposes, the amplitude of the microwave field $H_{\mu
w}\ll H$, so all the field effects should be regarded as coming from
the dc applied field.  Measurements were taken at several temperatures
in the range from 60 to 95 K, resulting in reduced temperature ranges
above $T/T_{c}=$ 0.55.  Data have been collected either in
zero-field-cooled, field-cooled, and on direct and reverse
field-sweeping.  Only at very low signal level (corresponding to low
fields) we observed a detectable hysteresis, which however affects
only a small part of the overall measured impedance variation (less
than 0.05 \% of $\rho_{1}$ in the normal state).  We briefly discuss
this effect in the next section, but we anticipate that it is
irrelevant for the purposes of this paper.  As a matter of fact, we
found that in the entire temperature range explored the complex
response was reversible within our sensitivity for fields above 10 mT.
\section{Experimental results}
\label{results}
\begin{figure}
  \includegraphics[width=0.4\textwidth]{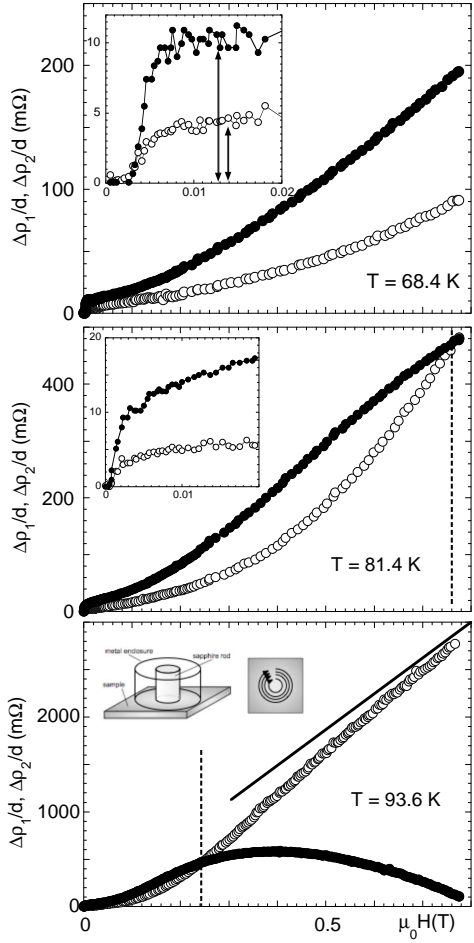}
\caption{Typical measurements of the field induced change of the
complex resistivity $\Delta\tilde\rho$ at 47.7 GHz at various
temperatures.  Note the very different vertical scales of the panels.
For comparison, $\rho_{1}/d \approx$ 10 $\Omega$ above $T_{c}$.  Open
circles: real part $\Delta\rho_{1}/d$.  Full dots: imaginary part
$\Delta\rho_{2}/d$.  Vertical dashed lines mark the crossover field
from elastic ($\Delta\rho_{2} > \Delta\rho_{1}$) to dissipative
($\Delta\rho_{1} > \Delta\rho_{2}$) response with increasing
temperature and field.  Continuous line in the lower panel indicates a
$\propto H$ law, typical of flux flow or rigid pinning.  As can be
seen, even at 93.6 K these regime are not achieved.  Upper and mid insets: low field
typical step-like variation of the complex resistivity, ascribed to
weak links related phenomena. Lower panel inset: sketch of the 
experimental configuration and of the pattern of microwave currents on 
the sample.}
\label{rampe}
\end{figure}

We focus here on the data taken on one particular TBCCO sample, as
representative of the behavior observed in different films prepared as
described in Sec.\ref{exp}.\\
Figure \ref{rampe} shows typical data for the
complex resistivity shift at low, intermediate and high temperatures.\\
First, let us briefly discuss the low field behaviour.  From the data
shown in the insets of Fig. \ref{rampe}, it is seen that a small
feature (note the expanded vertical scale, and compare to the scales
for measurements at higher $T$) is present below 10 mT. This feature
is hysteretic, in both real and imaginary part, at low temperatures
(below 70 K) and reversible above, with the hysteresis (when present)
depending on magnetic hystory.  This behaviour and the saturating
field dependence after the initial steep increase are strongly
reminiscent of the response of a network of weak links, due to e.g.
granularity of the film under study.  The distinction between all the
mechanisms involved in presence of weak links
\cite{giura,halbritter,gurevich} is not within the purposes of the
present paper.  However, as an indication for future work, it is
interesting to note that in thin TBCCO films of the same origin as
ours a nonlinear surface resistance at 8.5 GHz has been ascribed to AJ
fluxon motion, and was found to saturate at (microwave) fields as low
as $\mu_{0}H_{0}\sim$ 1 mT \cite{gaganidze}, similar to our range of
the dc fields $\sim$ 1-10 mT at which the low field linear response
saturates.  We then argue that, in our measurements, a small part of
the response is probably due to JF or AJF dynamics, even if this does
not appear to be the main mechanism driving the overall measured microwave
response.  In the following we will concentrate on the major part of
the microwave response, above the plateau of the low-field signal.  In
particular, we will subtract this small contribution from the
remaining of the curve, that we will treat in the framework of the
dynamics of Abrikosov vortices. As a consequence, we will identify our 
measured $\Delta\tilde\rho$ with $\tilde\rho_{vm}$.\\
Looking at Figure \ref{rampe}, several features
can be highlighted.  First, the magnetic field dependence is clearly
superlinear both in $\Delta\rho_{1}$ and $\Delta\rho_{2}$ in a wide
range of temperatures, showing no $\tilde\rho\propto H$.  Second,
while $\Delta\rho_{1}$ keeps increasing at all temperatures and
fields, by approaching $T_{c}$ $\Delta\rho_{2}$ changes shape, and at a
certain field begins to decrease.  In particular, it is seen that the
complex resistivity changes several times its behaviour: at low
temperatures and fields $\Delta\rho_{1}\sim\Delta\rho_{2}$; in an
intermediate region one has $\Delta\rho_{2}>\Delta\rho_{1}$ (elastic
response); by increasing field and temperature we observe a crossover from
predominantly elastic (Fig. \ref{rampe}, upper panel) to predominantly dissipative
(Fig. \ref{rampe}, lower panel) behaviour, marked by the crossing point of
$\Delta\rho_{1}$ and $\Delta\rho_{2}$.\\
On qualitative grounds, several remarks can be done.  A large,
positive reactive response can be due to the elastic response of
vortex lines.  In principle, it can be due to single-vortex pinning,
collective pinning, elasticity of the lattice, and also to some more
exotic origins such as surface pinning \cite{placais}.  The collapse of the reactive
response indicates that the vortex lines no more respond to some
recalling force (whatevere the origin), as reasonably expected when
the temperature increases sufficiently. 
Thus, the qualitative features seem to indicate that the changes in
$\Delta\tilde\rho$ are dominated by vortex motion.
However, a quantitative analysis is not straightforward, as we point 
out in the next section.
\section{Discussion}
\label{disc}
\begin{figure}
  \includegraphics[width=0.4\textwidth]{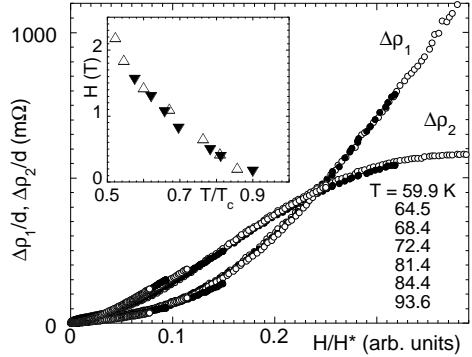}
\caption{Field scaling of the complex resistivity $\Delta\tilde\rho$.
With a temperature-dependent field, all pairs of curves collapse onto
a single curve.  In this case, we have subtracted the small
junction-related signal from the total response.  Inset: scaling field
(full triangles) compared to the melting field $H_{m}$ as obtained from dc
$I-V$ measurements in a TBCCO single crystal \cite{ammor} (scaled by 
a factor 2.2, open triangles).}
\label{scaling}
\end{figure}

\begin{figure}
  \includegraphics[width=0.4\textwidth]{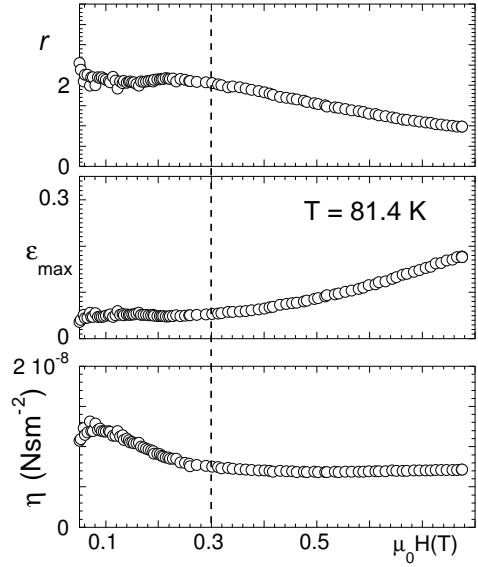}
\caption{Illustration of the estimate of the maximum creep factor, the vortex
viscosity and the pinning frequency at $T=$ 81 K.
Upper panel: ratio $r$ as defined in
Eq.\ref{creep}. Above the dashed line, $r\simeq\frac{\nu_{p}}{\nu}$.
Mid panel: maximum creep
factor as given by Eq.\ref{ineq}.  It is seen that the maximum creep factor
remains small at every field.
Lower panel: calculated vortex viscosity.  The procedure is
selfconsistent if $\eta$ is not field dependent.  Here, for fields
above 0.3 T, marked by the dashed line, the procedure yields reliable
estimates of $\eta$ and $\nu_{p}$.}
\label{param}
\end{figure}

First, we notice that the data are not in agreement with single, rigid
vortex pinning.  In this case, the field dependence of the response
should depend linearly on the vortex density \cite{gittle,cc,brandt},
that is one should have $\Delta\tilde\rho\propto B$, which is clearly
not the case even at high temperatures (Fig.\ref{rampe}, lower panel).
One could invoke nonuniform magnetic field penetration, but this seems
unlikely because (1) the area probed by the microwave field is of
diameter $\sim$ 2 mm, centered on a 10 mm $\times$ 10 mm film, (2) the
extremely high demagnetization factor makes vortex penetration nearly
coincident with the application of the field, (3) we did not observe
any difference between field-cooled, zero-field-cooled and swept-field
measurements (apart the tiny effect discussed above) and (4) the
nonproportionality to $H$ extends up to high temperatures.\\
A starting point for the analysis of the data is the general property exemplified in
Fig.\ref{scaling}: there, we show that our data for
$\Delta\rho_{1}$ and $\Delta\rho_{2}$ obey an experimental scaling
law, $\Delta \tilde \rho(H,T)=\Delta \tilde \rho[H/H^{*}(T)]$, where
the temperature dependent scaling field $H^{*}(T)$ is reported in the
inset.  The temperature dependence of $H^{*}(T)$ is compared to the temperature
dependence of the melting field $H_{m}(T)$ as determined by $I-V$
measurements in Tl:2212 crystals \cite{ammor}.\\
While this is a purely empirical scaling, nevertheless the quality of
the collapse is remarkable. It should be mentioned that, at low 
reduced fields and low temperatures, the quality deteriorates, indicating 
eventually a breakdown of such a simple construction. However, the scaling suggests that
the dynamics is mainly governed by one field scale at intermediate 
and high fields, while only at low fields additional mechanisms can play a 
role. The overall behaviour of $\Delta \tilde \rho[H/H^{*}(T)]$ 
indicates a rather sharp increase of the vortex mobility around the 
characteristic field $H^{*}$, as indicated by the increase of 
$\Delta\rho_{1}(H)$. Since the temperature 
dependence of the scaling field tracks the melting field, we speculate 
that the main features of the response are dictated by some kind of vortex 
transformation.\\
To substantiate the above suggestions and speculations, we come back
to the interpretation of the data by means of Eq.\ref{ccb}. One of 
the main difficulties is to get information on the role of elastic 
response, as given by $\nu_{p}$, and of thermal depinning, as given 
by $\epsilon$ and $\nu_{0}$. It is useful to introduce the ratio $r$ 
defined in terms of measured quantities and related to theoretical 
parameters as:
\begin{equation}
\label{creep}
r=\frac{\Delta\rho_{2}}{\Delta\rho_{1}}=
\frac{\nu_{0}}{\nu}\frac{1-\epsilon}{1+\left(\frac{\nu_{0}}{\nu}\right)^{2}\epsilon}
\end{equation}
With this definition, it can be shown on general grounds that, whatever the value of the 
characteristic frequency, one has
\begin{equation}
\label{ineq}
\epsilon <\epsilon_{max}=1+2r^{2}-2r\sqrt{r^{2}+1}
\end{equation}
First, we comment on the intermediate field region, where
$\Delta\rho_{2}>\Delta\rho_{1}$.  On the basis of Eq.\ref{ineq}, in
this region $\epsilon\ll$1, and $r\simeq\frac{\nu_{p}}{\nu}$.  Thus,
$\eta$ can be calculated from Eq.\ref{GR}.  To be consistent, the
procedure must yield a field-independent vortex viscosity.  In
Fig.\ref{param} we show the result of the procedure at an intermediate
temperature $T$=81.4 K. Above $\approx$0.3 T we get a field
independent $\eta$, indicating consistency of the procedure above 
that field.  In this region,
it turns out that the deviations from $\Delta\tilde\rho\propto H$ have
to be ascribed to a field dependence of $k_{p}$.  In the same Figure
we report the calculated $r$.  In the field region where the procedure
of the extraction of the vortex parameters is meaningful (i.e., where
$\eta$ is constant with the field), we observe a decrease of
$r\simeq\nu_{p}$ with the field, on which we will come back later.\\
At low $T$ and $H$ the calculated vortex viscosity is no more constant
with the field, so that the procedure is not selfconsistent: in this
case the interplay between $\nu_{0}$, $\nu_{p}$ and $\epsilon$ in
Eq.\ref{creep} does not allow for a direct calculation of vortex
parameters (unless multifrequency measurements can be done
\cite{sartiPhC,tsuchiya,belk,powell}).  However, it is still possible
to obtain some information by Eq.\ref{ineq}, in terms of an upper
bound for the creep factor.  We focus momentarily on the region of
moderate fields (0.1 $\div$ 0.3 T) to have an estimate of the ``zero
field'' creep factor (upper bound) without possible artificial
contributions from the small, weak-link related, anomaly.  The
calculation of the upper bound for $\epsilon$ yields small
$\epsilon_{max}$ increasing from 0.015 at 60 K to 0.04 at 93.6 K. We
stress again that this is only an upper bound, and nothing can be
determined on, e.g., the field dependence of the true creep factor.
Nevertheless, the upper bound for $\epsilon$ gives a lower bound for the activation
energies.  Using $\epsilon_{max}\approx e^{-U_{min}/k_{B}T}$, we find a
nearly temperature independent $U_{min}= (18 \pm 4)$ meV (in the field
region between 0.1 and 0.3 T).  It is then possible to say that the
potential seen by the flux lines in our TBCCO film has barriers of
height that might extend down to $\sim$ 20 meV in the temperature range
explored at low fields.\\
Now we comment on the obtained results, and we propose a general 
framework for our findings.\\
Low activation energies have been already reported in YBCO films as
probed by microwave measurements \cite{belk,powell}.  Remarkably, the
activation energies reported in YBCO \cite{belk,powell} are compatible
with the lower bound here reported for TBCCO. In both cases, the small
activation energies must compare with finite critical currents and
vanishing resistivity in the same field and temperature range.  It is
then natural to give an interpretation in terms of a very ``rough''
potential for flux lines, with deep wells effective for large
displacements (dc currents), and many small wells which determine the
effective potential barrier at high driving frequencies.  As discussed
by others \cite{belk,powell}, such small activation energies can be
taken as a manifestation of a ``glassy'' state, where it is not
possible to push further the applicability of the CCB model, and
distributions of energy barriers have to be taken into account.
However, with respect to YBCO a significant difference in TBCCO
appears with increasing field, indicated by the increase of vortex
mobility and the drop of the imaginary part.  In this case, using the
CCB model a field dependent pinning frequency is obtained which is not
observed in YBCO in the same field range.  Even if this feature could
be determined by several effects, we remark that the procedure to
obtain the vortex parameters is mostly consistent in the region where
the mobility increases (things that should not happen if creep effects
were significant).  Thus, we are led to assign the observed field
dependence of the global parameter $\nu_{p}$ to a field dependent
$k_{p}$.  The drop in $\nu_{p}$ can then find an explanation in
terms of loss of vortex rigidity: in the strongly anisotropic compound
TBCCO, above a characteristic field the vortex can decouple or, at
least, become weakly correlated along the $c$ axis.  In this case the
elastic moduli decrease with the field \cite{koshelev}, and the
elastic response for small displacements becomes dominated by the
``weak spring'' given by the reduced elasticity of the flux line.  In
this frame, the scaling that we find in our data is consistent: the
dynamics is dominated by the loss of rigidity, apart the low field/low
temperature region.  The crossover between the ``mainly rigid'' and
``mainly uncorrelated'' regimes is given by the field scale $H^{*}$,
that in fact behaves as a function of the temperature as the vortex
melting field \cite{ammor}.\\
While it is clear that many details still need to be worked out, the 
present results give an account for the complex microwave response at high 
driving frequencies in TBCCO in the vortex state.
\section{Conclusion}
\label{conc}
We have presented data for the microwave resistivity in
Tl$_{2}$Ba$_{2}$CaCu$_{2}$O$_{8+x}$ thin films as a function of the
temperature and magnetic field.  Apart the very low fields region,
where there is a possible contribution from granularity-driven
processes, most of the response seems to be dominated by vortex
motion.  The field-dependence of the complex resistivity brings
indication for strong elastic response, indicated by
$\Delta\rho_{2}>\Delta\rho_{1}$.  We find a noteworthy scaling
behaviour of the complex resistivity, where the dynamics seem to be
dictated by a single field scale $H^{*}(T)$.  $H^{*}(T)$ reproduces
the temperature dependence of the vortex melting field in TBCCO
crystals, suggesting that the vortex dynamics, even at our high
measuring frequencies, is dictated by some vortex transformation.  The
analysis by means of the Coffey-Clem-Brandt model indicates a dynamics
driven by the loss of correlation of the flux lines along the $c$
axis, as expected in layered superconductors with increasing field and
temperature.  We also estimate a lower bound for the activation
energies for thermal depinning $U_{min}$.  On the basis of the failure
of the CCB model at low fields and temperatures, together with the low
estimates for $U_{min}$, we suggest that at low fields the potential
landscape is very rough with many low local barriers, that can give
rise to a peculiar vortex dynamics,
similarly to observations in YBCO. The larger anisotropy in TBCCO is 
suggested to be responsible for the increased mobility above a characteristic field, above 
which the CCB approach is self consistent and vortex parameters can 
be estimated.\\

%


\begin{thebibliography}{}

\bibitem{tomasch} W. J. Tomasch, H. A. Blackstead, S. T. Ruggiero, P.
J. McGinn, J. R. Clem, K. Shen, J. W. Weber, and D. Boyne, {\it Phys.
Rev.  B}, \textbf{37} 9864 (1988)

\bibitem{golos} M. Golosovsky, M. Tsindlekht, D. Davidov, {\it 
Supercond. Sci. Technol.} {\bf 9}, 1 (1996), and references therein

\bibitem{gittle} J. I. Gittleman and B. Rosenblum, {\it Phys. Rev. Lett.} 
{\bf 16}, 734 (1966)

\bibitem{cc} M.W. Coffey and J.R. Clem, {\it Phys.  Rev.  Lett.} 
{\bf 67}, 386 (1991)

\bibitem{brandt} E. H. Brandt, {\it Phys.  Rev.  Lett.} 
{\bf 67}, 2219 (1991)

\bibitem{revenaz} S. Revenaz, D. E. Oates, D. Labb\'e$-$Lavigne, G. 
Dresselhaus, and M. S. Dresselhaus, {\it Phys.  Rev.  B}
{\bf 50}, 1178 (1994)

\bibitem{sartiPhC} S. Sarti, E. Silva, C. Amabile, R.Fastampa, 
M.Giura, {\it Physica C} {\bf 404}, 330 (2004)

\bibitem{marcon} R.Marcon, R.Fastampa, M.Giura, E.Silva, {\it Phys.  Rev. B} 
{\bf 43}, 2940 (1991)

\bibitem{tsuchiya}
Y. Tsuchiya, K. Iwaya, K. Kinoshita, T. Hanaguri, H. Kitano, A. 
Maeda, K. Shibata, T. Nishizaki, and N. Kobayashi, {\it Phys. Rev. B} 
{\bf 63}, 184517 (2001)

\bibitem{pompeoJPCS}
N. Pompeo, L. Muzzi, S. Sarti, R. Marcon, R.
Fastampa, M. Giura, M. Boffa, M.C. Cucolo, A.M. Cucolo, C.
Camerlingo, E. Silva, {\it J. Phys.  Chem.  of Solids} {\bf 67},
460 (2006)

\bibitem{ffh} D. S. Fisher, M. P. A. Fisher, D. A. Huse, {\it Phys. Rev. B} {\bf 
43}, 130 (1991)

\bibitem{blatter} G. Blatter, M. V. Feigel'man, V. B. Geshkenbein, A. 
I. Larkin, V. M. Vinokur, {\it Rev. Mod. Phys.} {\bf 
66}, 115 (1994)

\bibitem{wu}
D. H. Wu, J. C. Booth, and S. M. Anlage, {\it Phys.  Rev.  Lett.} {\bf 
75}, 525 (1995)

\bibitem{giura} R. Marcon, R. Fastampa, M. Giura, C. Matacotta, {\it
Phys.  Rev.  B} {\bf 39}, 2796 (1989); M. Giura, R. Marcon, R.
Fastampa, {\it Phys.  Rev.  B} {\bf 40}, 4437 (1989); M. Giura, R.
Fastampa, R. Marcon, E. Silva, {\it Phys.  Rev.  B} {\bf 42}, 6228
(1990)

\bibitem{halbritter} J. Halbritter, {\it J. Supercond.} {\bf 8}, 691
(1995)

\bibitem{gurevich} A. Gurevich, {\it Phys.  Rev.  B} {\bf 46}, 3187 (1992); {\it
Phys.  Rev.  B} {\bf 65}, 214531 (2002)

\bibitem{tsindlekht} M.I. Tsindlekht, E.B. Sonin, M.A. Golosovsky, D.
Davidov, X. Castel, M. Guilloux-Viry, A. Perrin, {\it Phys.  Rev.  B}
{\bf 61}, 1596 (2000)

\bibitem{gaganidze} E. Gaganidze, R. Heidinger, J. Halbritter, A. 
Shevchun, M. Trunin, H. Schneidewind, {\it J. Appl. Phys.}
{\bf 93}, 4049 (2003)

\bibitem{schneidewind} 
H. Schneidewind, M. Manzel, G. Bruchlos, and K. Kirsch, {\it
Supercond.  Sci.  Technol.} {\bf 14}, 200 (2001); H. Schneidewind,
M. Zeisberger, H. Bruchlos, M. Manzel, T. Kaiser, {\it Institute of
Physics Conf.  Series} {\bf 167}, 383 (2000)

\bibitem{silvaMST} E. Silva, A. Lezzerini, M. Lanucara, S. Sarti and 
R. Marcon, {\it Meas.  Sci.  Technol.} {\bf 9}, 275 (1998)

\bibitem{pompeoJSsub} N. Pompeo, R. Marcon, E. Silva, submitted to {\it J. 
Supercond.} (2006)

\bibitem{sridhar} S. Sridhar, {\it J. Appl.  Phys.} {\bf 63}, 159
(1988)

\bibitem{silvaSUSTthin}
E. Silva, M. Lanucara, R. Marcon, {\it Supercond.  Sci.  Technol.} 
{\bf 9}, 934 (1996)

\bibitem{placais} B. Placais, P. Mathieu, Y. Simon, E. B. Sonin and 
K. B. Traito, {\it Phys.  Rev.  B}
{\bf 54}, 13083 (1996)

\bibitem{ammor} L. Ammor, J. C. Soret, A. Smina, V. Ta Phuoc, A.
Ruyter, A. Wahl, B. Martinie, J. Lecomte and Ch.  Simon, {\it Physica
C} {\bf 282} 1983, (1997)

\bibitem{belk} N. Belk, D. E. Oates, D. A. Feld, G. 
Dresselhaus, and M. S. Dresselhaus, {\it Phys.  Rev.  B}
{\bf 53}, 3459 (1996)

\bibitem{powell} J. R. Powell, A. Porch, R. G. Humphreys, F. 
Wellh\"ofer, M. J. Lancaster, C. E. Gough, {\it Phys.  Rev.  B}
{\bf 57}, 5474 (1998)

\bibitem{koshelev} A. E. Koshelev and P. H. Kes, {\it Phys.  Rev.  B}
{\bf 48}, 6539 (1993)

\end{thebibliography}
\end{document}